\documentclass{PoS}
\usepackage{epsfig}

\title{Diphoton + jets at NLO}

\ShortTitle{Diphoton + jets at NLO}

\author{Thomas Gehrmann\\
        Institute for Theoretical Physics, University of Z\"urich, Winterthurerstrasse 190, CH-8057 Z\"urich, Switzerland \\
        E-mail: \email{thomas.gehrmann@uzh.ch}}

\author{\speaker{Nicolas Greiner}%
        \\
        Max Planck Institute for Physics, F\"ohringer Ring 6, D-80805 M\"unchen, Germany\\
        E-mail: \email{greiner@mpp.mpg.de}}
\author{Gudrun Heinrich\\
        Max Planck Institute for Physics, F\"ohringer Ring 6, D-80805 M\"unchen, Germany\\
        E-mail: \email{gudrun@mpp.mpg.de}}

\abstract{We present the next-to-leading order QCD corrections to the production of a photon pair in association
with one or two jets. This class of processes constitutes an important background for Higgs physics at the LHC.
For the one jet process we include photon fragmentation contributions and perform a comparison between different
photon isolation criteria and various isolation parameters. The two jet calculation has been performed using
a smooth cone isolation criterion. We find that the NLO QCD corrections are substantial and can lead to distortions
of differential distributions.}

\FullConference{11th International Symposium on Radiative Corrections (Applications of Quantum Field Theory to Phenomenology) \\
		 22-27 September 2013\\
		 Lumley Castle Hotel, Durham, UK}

\begin{document}

\section{Introduction}
The production of a photon pair at the LHC within the Standard Model is an important background for the Higgs 
boson observed at the LHC~\cite{higgsLHC}. One of the most important decay channels is the decay into a photon
pair~\cite{haa,cmshaa,atlashaa}, due to its clean signature. The Standard Model background,
the production of two photons without an intermediate Higgs, is experimentally determined by means of
sideband subtraction. However, for a more precise determination of the Higgs parameters, this estimation of
the background might not be precise enough. Therefore it is crucial to have a reliable theoretical prediction
for this class of processes. The production of two photons is known fully differentially at NLO~\cite{diphotonNLO},
including gluon initiated subprocesses~\cite{Bern:2002jx} and resummation effects~\cite{Balazs:1999yf,Balazs:2006cc}. 
Recently also the NNLO result has become available~\cite{ggaannlo}.\\
The presence of extra jets allows for a better control of the backgrounds and for a distinction of different
Higgs production processes. In particular, the additional production of two jets allows to probe the vector
boson fusion production mechanism. Therefore the QCD corrections for processes with the presence of extra jets 
are an important ingredient for reliable predictions in the diphoton plus jets channel. 
The production of a photon pair in association with one
jet is known at NLO~\cite{nagy} using a smooth cone isolation criterion~\cite{frixione}.
We present an NLO QCD  calculation for diphoton plus one jet where 
for the first time different photon isolation criteria and different
isolation parameters are compared, and we present the first NLO calculation of diphoton plus two-jet production.

\section{Photon isolation and photon fragmentation}
A final state photon can either originate from the hard scattering matrix elements which means it is perturbatively
accessible, or it can originate from the fragmentation of hadrons into photons. The latter production mechanism is 
of non-perturbative nature and therefore can not be described using a fixed order calculation.
Only the production of photons from the hard interaction can be computed 
within perturbation theory from first principles, while the production of photons in hadronization
and hadron decays can only be modeled, thereby introducing a dependence on rather poorly known fragmentation parameters.\\
The characteristics of the photon typically depend on its origin. Secondary photons, 
i.e. photonns stemming from hadronic decays, are usually
inside a jet cone formed by the hadrons. 
Photons coming from the hard interaction tend to be separated from hadronic
jets. In order to distinguish the two types, one applies photon isolation cuts, 
which limit the hadronic activity around a photon candidate. 
However, a veto on all hadronic activity around the 
photon direction would result in a suppression of soft gluon radiation in part of the final state 
phase space, thereby violating infrared safety of the observables. Consequently, all 
photon isolation prescriptions must admit some residual amount of hadronic activity around the photon direction.
In the following we focus on two types of photon isolation. One is the cone-based isolation which is most commonly
used at hadron collider experiments. 
In this procedure, the photon candidate is identified (prior to the jet clustering) 
from its electromagnetic signature, and its momentum direction (described by 
transverse energy $E_{T,\gamma}$, rapidity $\eta_\gamma$ and polar angle $\phi_\gamma$) 
is determined. Around this momentum direction, a cone of radius 
$R_\gamma$ in rapidity $\eta$ and polar angle $\phi$ is defined. Inside this cone, the hadronic 
transverse energy  $E_{T,{\rm had}}$ is measured. The photon is called isolated if 
$E_{T,{\rm had}}$ is below a certain threshold, defined either in absolute terms, or as 
a fraction of $E_{T,\gamma}$ 
The latter criterion then means that a photon candidate is considered as isolated if it fulfills the condition
\begin{equation}
E_{{\rm had,cone}}  \leq \epsilon_c \,p_{T}^{\gamma} \;,
\label{eq:cone-iso}
\end{equation}
inside the cone with radius $R$. The cone-based isolation allows events with a quark being collinear
to a photon. The theoretical predictions for cross sections 
 defined with this type of isolation must therefore take account of photon fragmentation contributions.\\
A different option is the smooth cone isolation~\cite{frixione}. It varies the 
 threshold on the hadronic energy inside the isolation cone with the 
 radial distance from the photon. It is described by the cone size $R_\gamma$, 
 a weight factor $n_\gamma$ and 
 an isolation parameter $\epsilon_\gamma$. With this criterion, one considers smaller 
 cones of radius $r_\gamma$ inside the $R_\gamma$-cone and calls the photon isolated 
 if the energy in any sub-cone does not exceed
 \begin{equation}
 E_{{\rm had, max}} (r_{\gamma}) = \epsilon_\gamma p_{T}^{\gamma} \left( \frac{1-\cos r_\gamma}
 {1-\cos R_\gamma}\right)^{n_\gamma}\;.
 \label{eq:frix}
 \end{equation}
 By construction, the smooth cone isolation does not admit any hard collinear quark-photon
 configurations, thereby allowing a full separation of 
 direct and secondary photon production, and consequently eliminating the need for 
 a photon fragmentation contribution in the theoretical description. 
 Despite its theoretical advantages, the smooth cone isolation was up to now 
 used in experimental studies of isolated photons
 only in a discretized approximation. Experimentally, owing to finite detector resolution, 
 the  implementation of this isolation criterion will 
 always require some minimal, nonzero value of $r_{\gamma}$, thereby 
 leaving potentially a residual collinear contribution. 

\section{Diphoton + one jet}
\label{sec:1j}
A detailed calculation of this process and a comparison between the two photon isolation types can be
found in~\cite{aa1jnlo}. The code that has been used to produce the presented results  has
been made public and can be downloaded from \texttt{https://gosam.hepforge.org/diphoton/}.\\

\subsection{Setup of the calculation}
Both the calculation of diphoton plus one jet as well as  with two jets, described
in section \ref{sec:2j}, are performed using the same setup.
For the generation of the tree level and real emission matrix elements we use MadGraph~\cite{mg4},
the regularization of infrared QCD singularities is handled by MadDipole~\cite{maddipole},
which makes use of the dipole formalism as developed in Ref.~\cite{Catani:1996vz}. For integration over
the phase space we used MadEvent~\cite{Maltoni:2002qb}. The routines for generating histograms and distributions
originate from the MadAnalysis package (see http://madgraph.hep.uiuc.edu).
All ingredients are generated and combined in a fully automated way.\\
The virtual amplitudes have been obtained using GoSam~\cite{gosam}, a public package for the automated
generation of one-loop amplitudes. It uses Qgraf~\cite{qgraf}, FORM~\cite{form} and Spinney~\cite{spinney}
for the generation of the amplitudes, which are optimized and written as a Fortran90 code with the use
of Haggies~\cite{haggies}. For the reduction of the one-loop amplitudes we used Samurai~\cite{samurai},
which performs a reduction at the integrand level based on the OPP-method~\cite{unitarity}.
As a rescue for numerically unstable points, the tensor reduction method of the Golem95 library~\cite{golem95} has
been used. The remaining master integrals are computed with either OneLoop~\cite{oneloop}, 
or Golem95~\cite{golem95}.\\
In the case where the fixed cone isolation is used, the real emission contribution contains 
infrared singularities related to the collinear emission of the photon off a final-state QCD parton.
They  need to be regularized by some kind of subtraction terms.
The corresponding integrated subtraction terms make this singularity apparent as they contain
an explicit pole term  when integrating over the unresolved one-particle phase space in
dimensional regularization. This pole is then absorbed into the fragmentation
functions.
To regulate these singularities we again make use of the dipole formalism 
as developed in Ref.\cite{Dittmaier:1999mb}
and implemented in the QED extension of MadDipole \cite{Gehrmann:2010ry}. This extension also
offers the framework for a straightforward implementation of fragmentation functions.

\subsection{Numerical Results}
For the results presented here we assumed as center of mass energy of $\sqrt{s}=8$\,TeV and we employed a
set of basic cuts:
$p_T^{\rm{jet}}>40$\,GeV, $p_T^{\gamma}>20$, $|\eta^{\gamma},\eta^j| \leq 2.5$, 
100\,GeV $\leq m_{\gamma\gamma} \leq $ 140\,GeV.
For the  jet clustering we used an anti-$k_T$ algorithm~\cite{Cacciari:2008gp} with a cone size
of $R=0.4$ provided by 
the FastJet package \cite{fastjet}.
We used an NLO pdf set from  NNPDF2.3 \cite{Ball:2012cx}, where the values for $\alpha_s$ at leading order 
and next-to-leading order are given by $\alpha_{s}(M_Z) = 0.119$.
For the photon fragmentation functions, we take set II of the 
parametrisations of Ref.~\cite{Bourhis:1997yu}.
Renormalization and factorization scales
$\mu_r$ and $\mu_F$ 
are dynamical scales and set to be equal, and we choose $\mu_0^2=\frac{1}{4}\,(m_{\gamma\gamma}^2+\sum_j p_{T,j}^2)$
for our central scale. When using cone based isolation, the fragmentation scale $\mu_f$ also enters, 
and we set it to be equal to the other scales.
To assess the effect of a jet veto, we present results for the inclusive and the exclusive case, where for
the latter we veto on a second jet by demanding $p_{T,jet2} < 30$ GeV.
For the photon isolation, we compare the Frixione isolation criterion with the fixed cone criterion
for several values of the photon energy fraction $z_c$ in the cone,
where $$z_c=\frac{|\vec{p}_{T,{\rm cone}}^{\rm{\,had}}|}{|\vec{p}_T^{\,\gamma}+\vec{p}_{T,{\rm cone}}^{\rm{\,had}}|}\;,$$ 
such that in the collinear limit,  $z_c$ is related to $\epsilon_c$ in   eq.~(\ref{eq:cone-iso}) by $z_c=\frac{\epsilon_c}{1+\epsilon_c}$. 
For the Frixione isolation criterion (see eq.~(\ref{eq:frix})), 
our default values are $R=0.4, n=1$ and $\epsilon=0.5$.
\\ 
 \FIGURE{
 \parbox{17.cm}{
 \epsfig{file=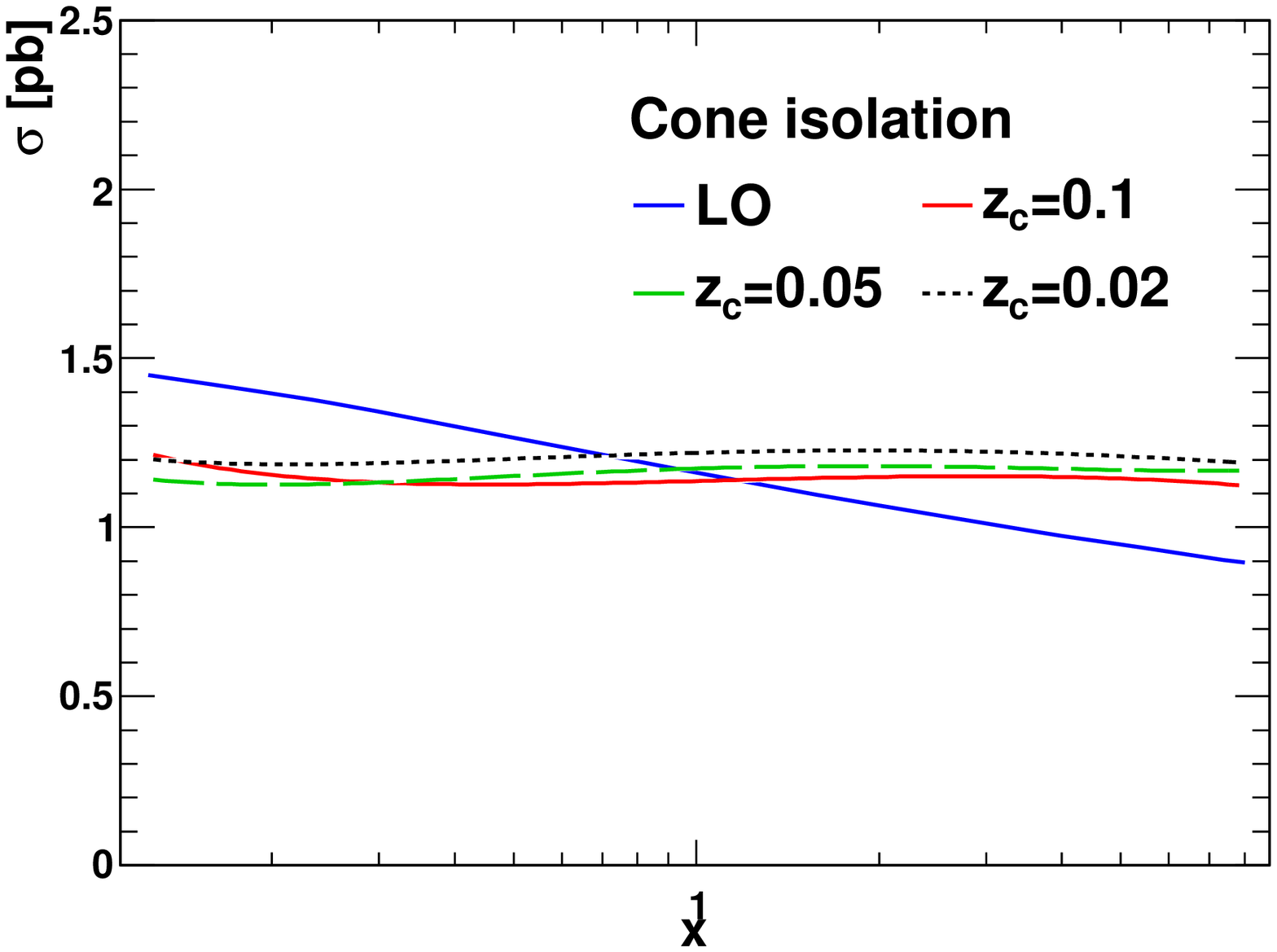,width=7.5cm}
 \epsfig{file=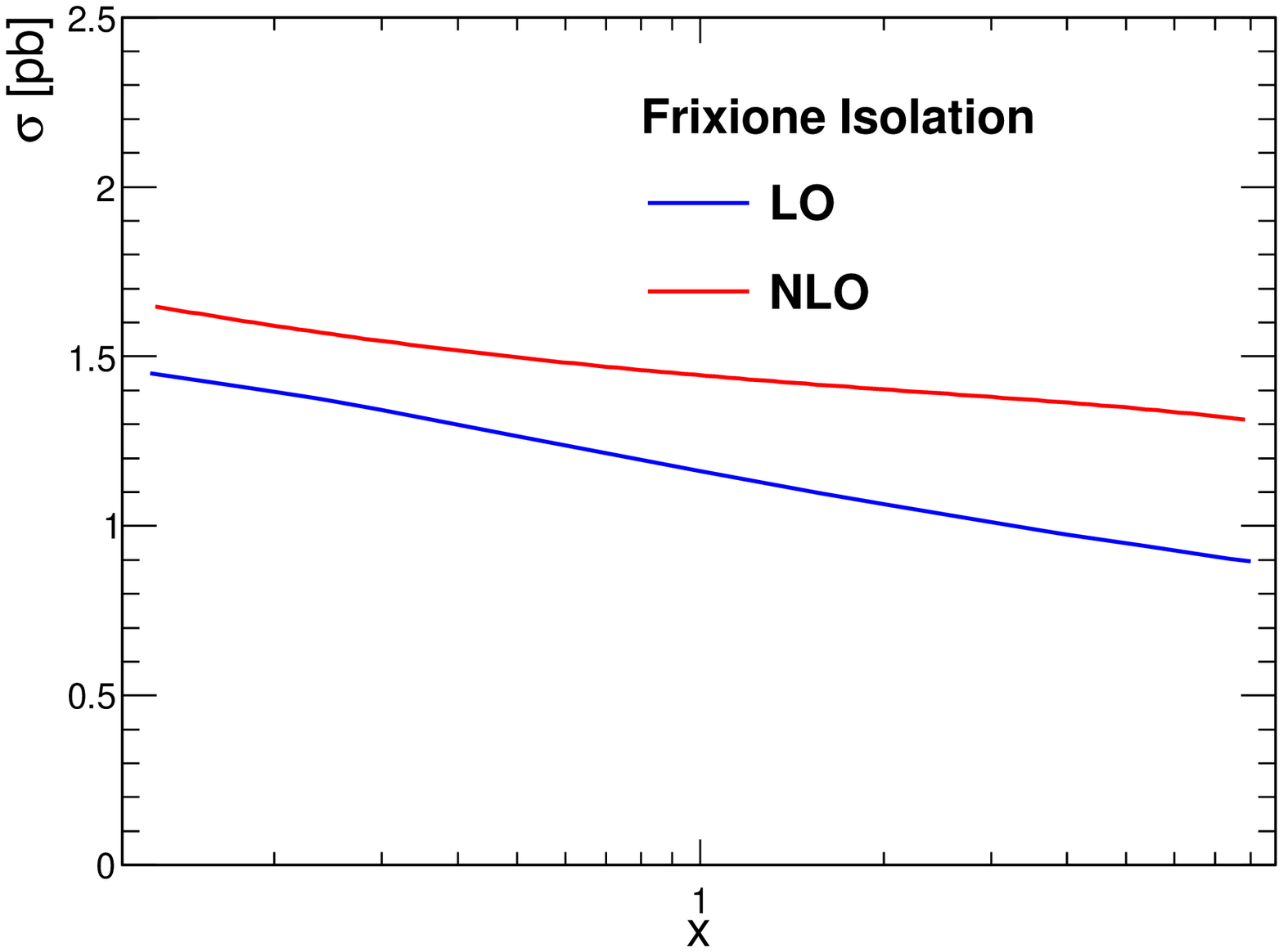,width=7.5cm}
 \caption{Behavior of the exclusive $\gamma\gamma$+jet  cross sections with different isolation
 prescriptions under scale variations, 
 $\mu=x\,\mu_0$,  $0.5\leq x \leq 2$, $\mu_0^2=\frac{1}{4}\,(m_{\gamma\gamma}^2+\sum_j p_{T,j}^2)$.
  \label{fig:scalevarexcl} }
 }
 }
We estimate the uncertainty stemming from renormalization, factorization, and, if applicable, fragmentation scale,
by varying these scales by a factor of two around the central value. For the exclusive case in Figure~\ref{fig:scalevarexcl}
we observe a clear reduction in the scale uncertainty whereas for the inclusive case in Figure~\ref{fig:scalevarincl}
this is not the case. The reason for this behavior is that real emission contribution dominates the total cross section, 
therefore leading to a tree level like scale dependence. For the exclusive case this contribution is
effectively cut off by the jet veto. 
However, the smallness of the scale dependence in this case has to be interpreted with care,
as this behavior depends on the jet veto and could be due to cancellations which might not be present in all distributions.
 
 \FIGURE{
 \parbox{17.cm}{
 \epsfig{file=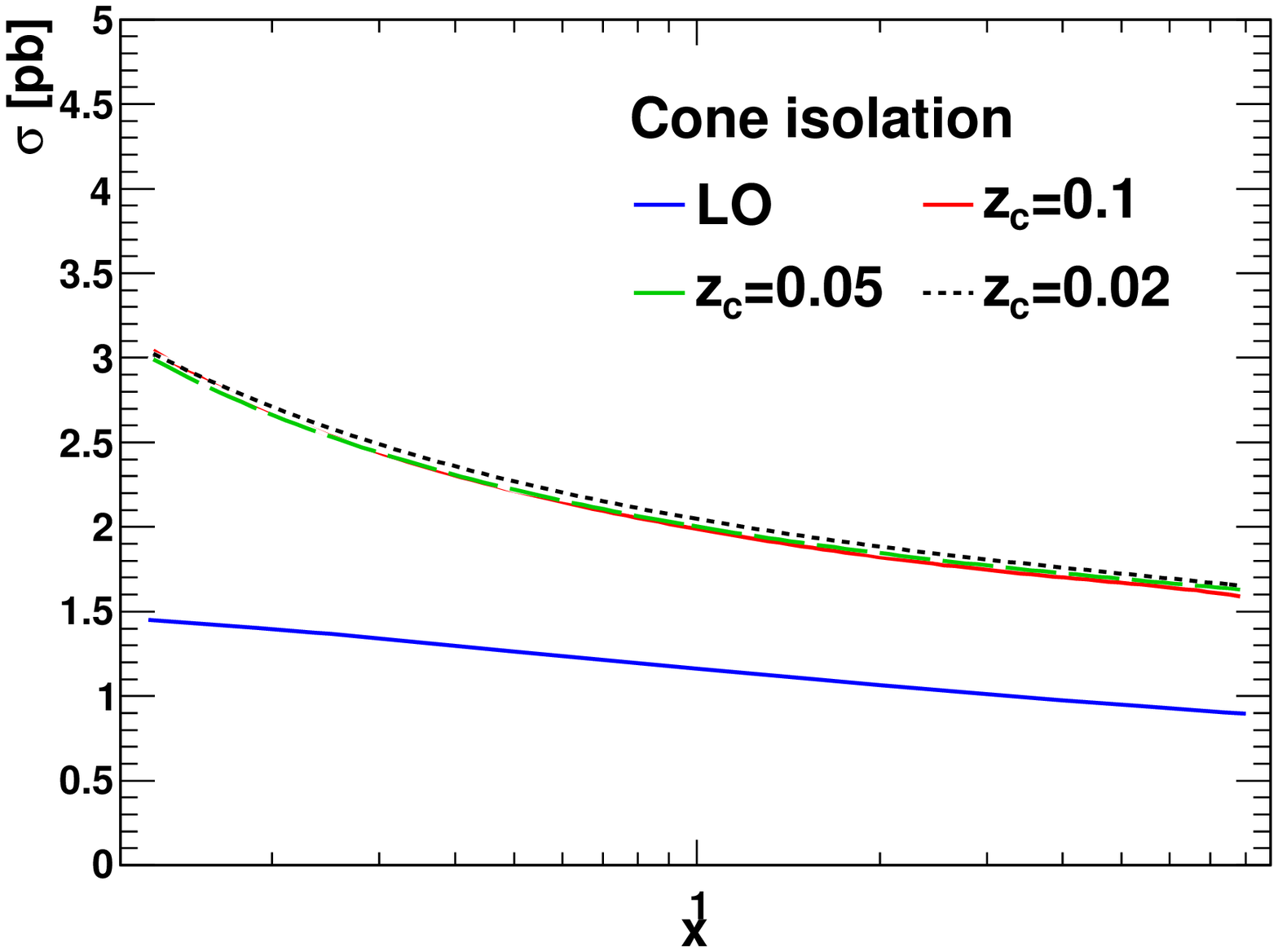,width=7.5cm}
 \epsfig{file=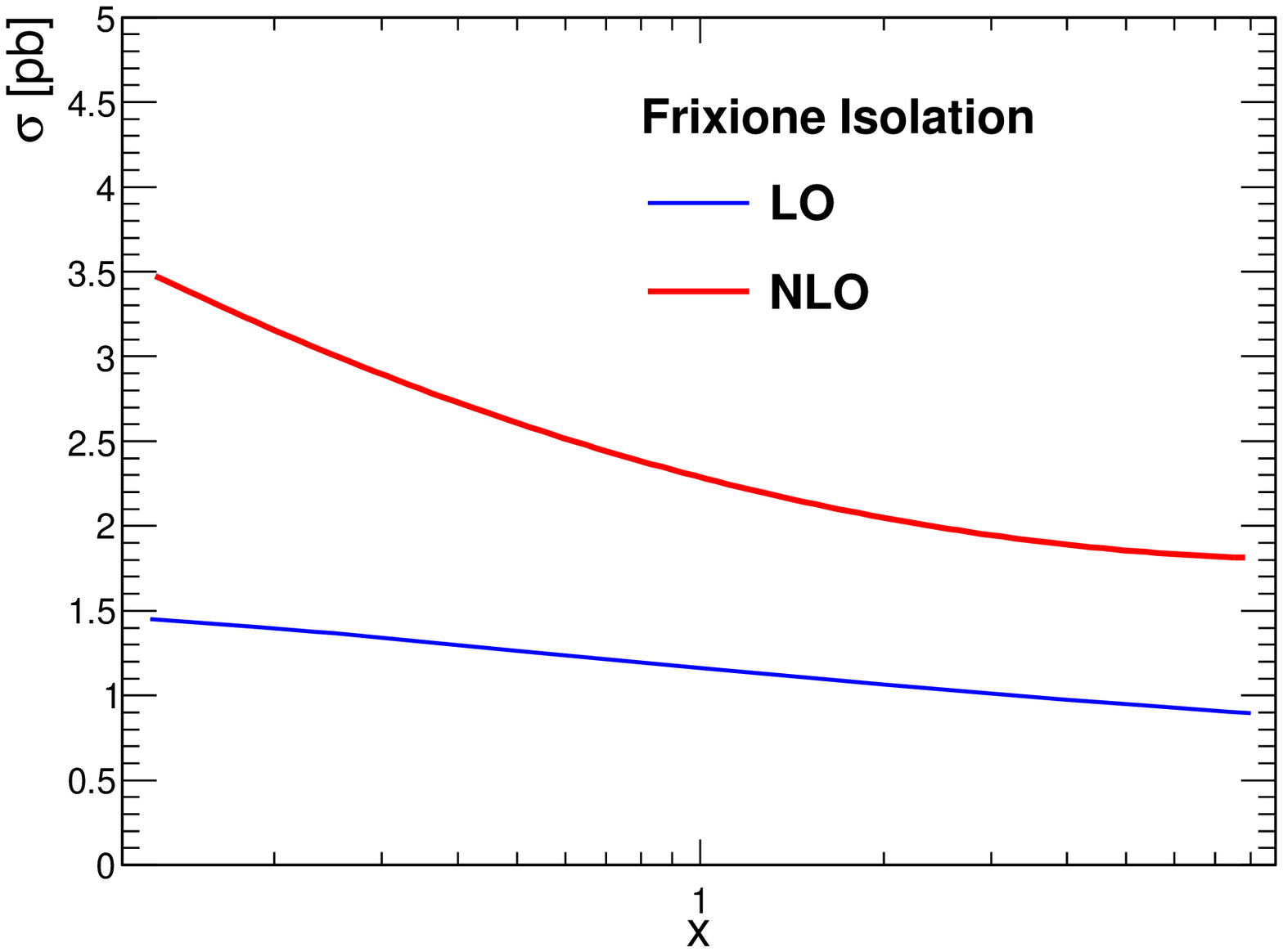,width=7.5cm}
 \caption{Behavior of the inclusive $\gamma\gamma$+jet+X cross sections with different isolation
 prescriptions under scale variations, 
 $\mu=x\,\mu_0$,  $0.5\leq x \leq 2$, $\mu_0^2=\frac{1}{4}\,(m_{\gamma\gamma}^2+\sum_j p_{T,j}^2)$.
  \label{fig:scalevarincl} }}
 }

\FIGURE{
 \parbox{17.cm}{
 \epsfig{file=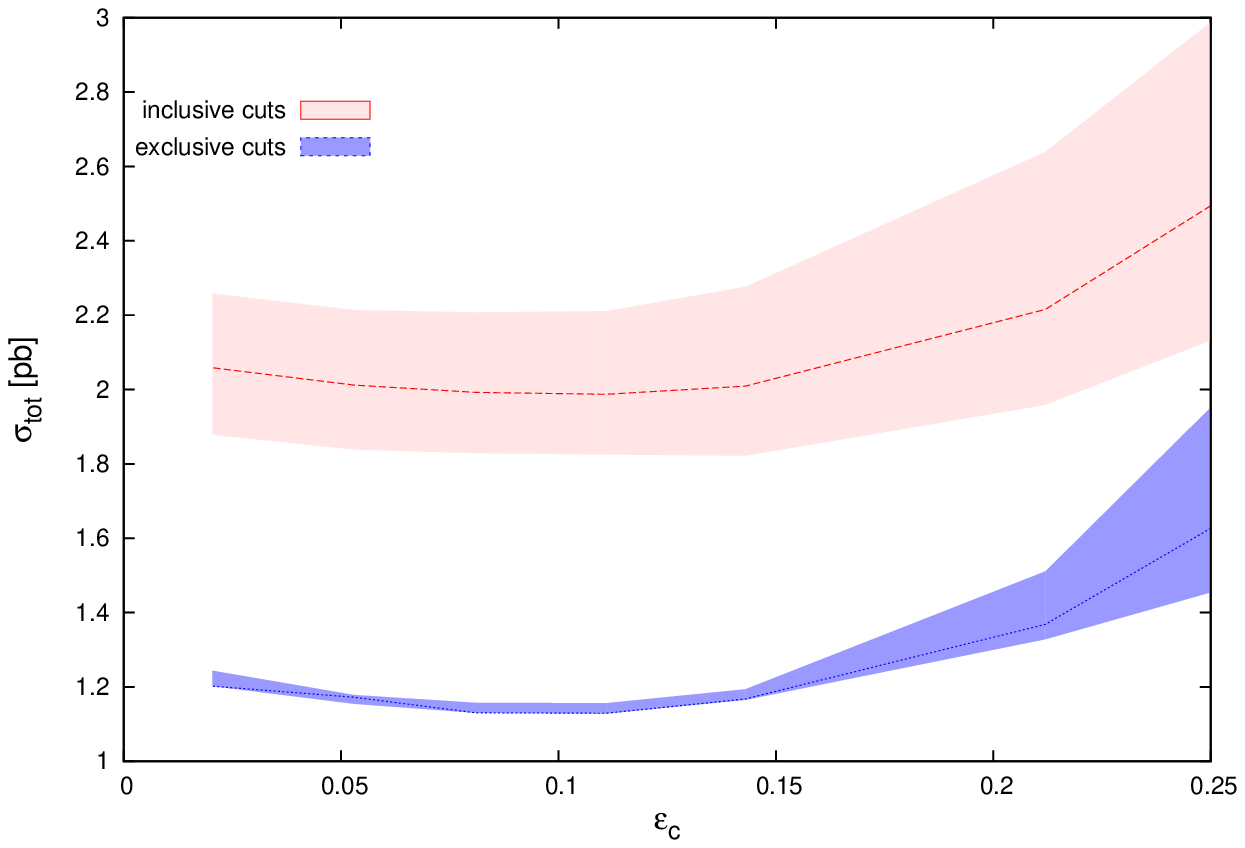,width=7.5cm}
 \epsfig{file=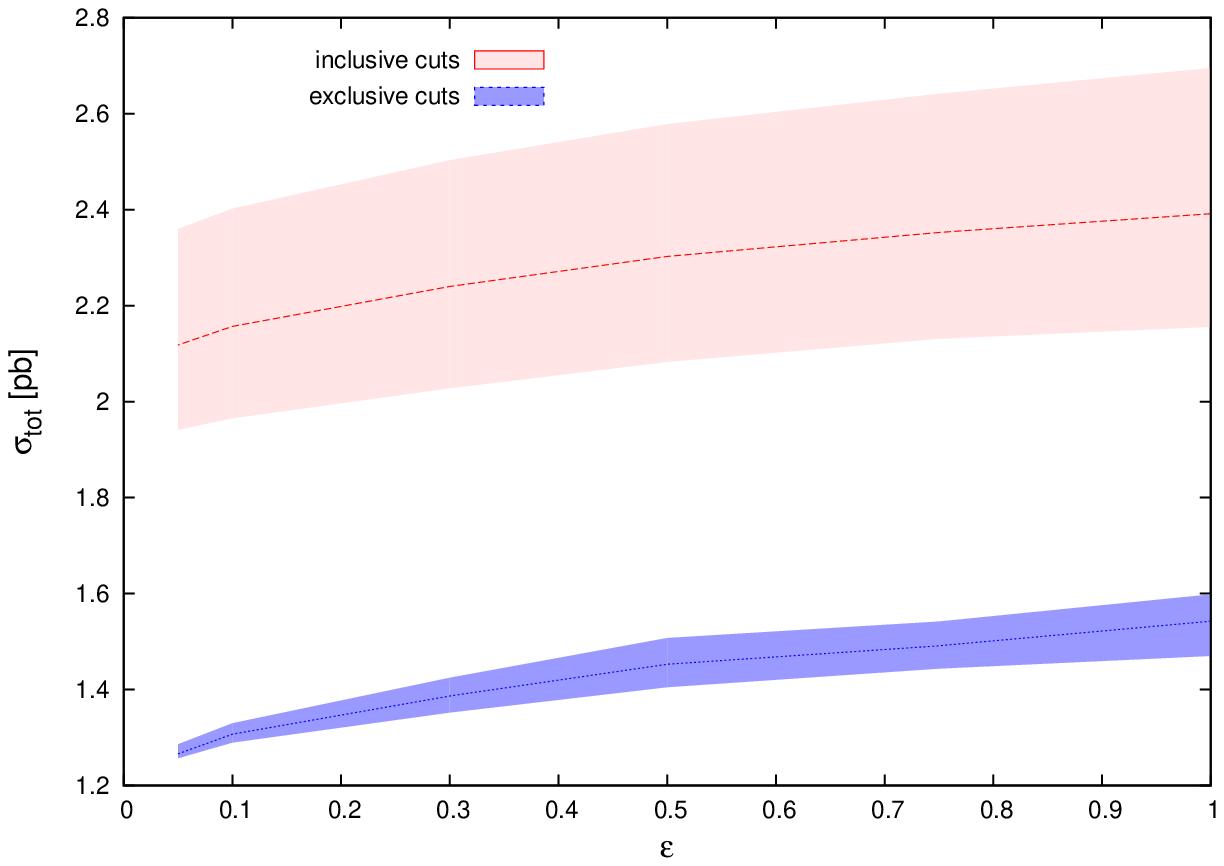,width=7.5cm}
 \caption{Dependence of the cross sections on the cone isolation parameters $\epsilon_c$ and $\epsilon$ respectively. 
 The bands correspond to scale variations $0.5\leq x \leq 2$.
  \label{fig:comp_z_Frix} }
 }
 }
 Figure \ref{fig:comp_z_Frix} shows how the scale variation bands vary as a function of the 
isolation parameters, for both the single-jet inclusive and the exclusive case. 
One observes that for both -- fixed cone isolation and Frixione 
isolation -- the inclusive case is dominated by the large scale dependence of the $\gamma\gamma+2$\,jets 
part of the real radiation which has an uncompensated  leading order scale dependence. 
  
\section{Diphoton + two jets}
\label{sec:2j}
For the calculation of a photon pair plus two jets we used the same tools and packages as described in section \ref{sec:1j}.
The calculation has first been described in~\cite{aa2j} to which we refer for a more detailed description.
\subsection{Numerical results}
\begin{figure}[h]
\begin{center}
\includegraphics[width=7.5cm]{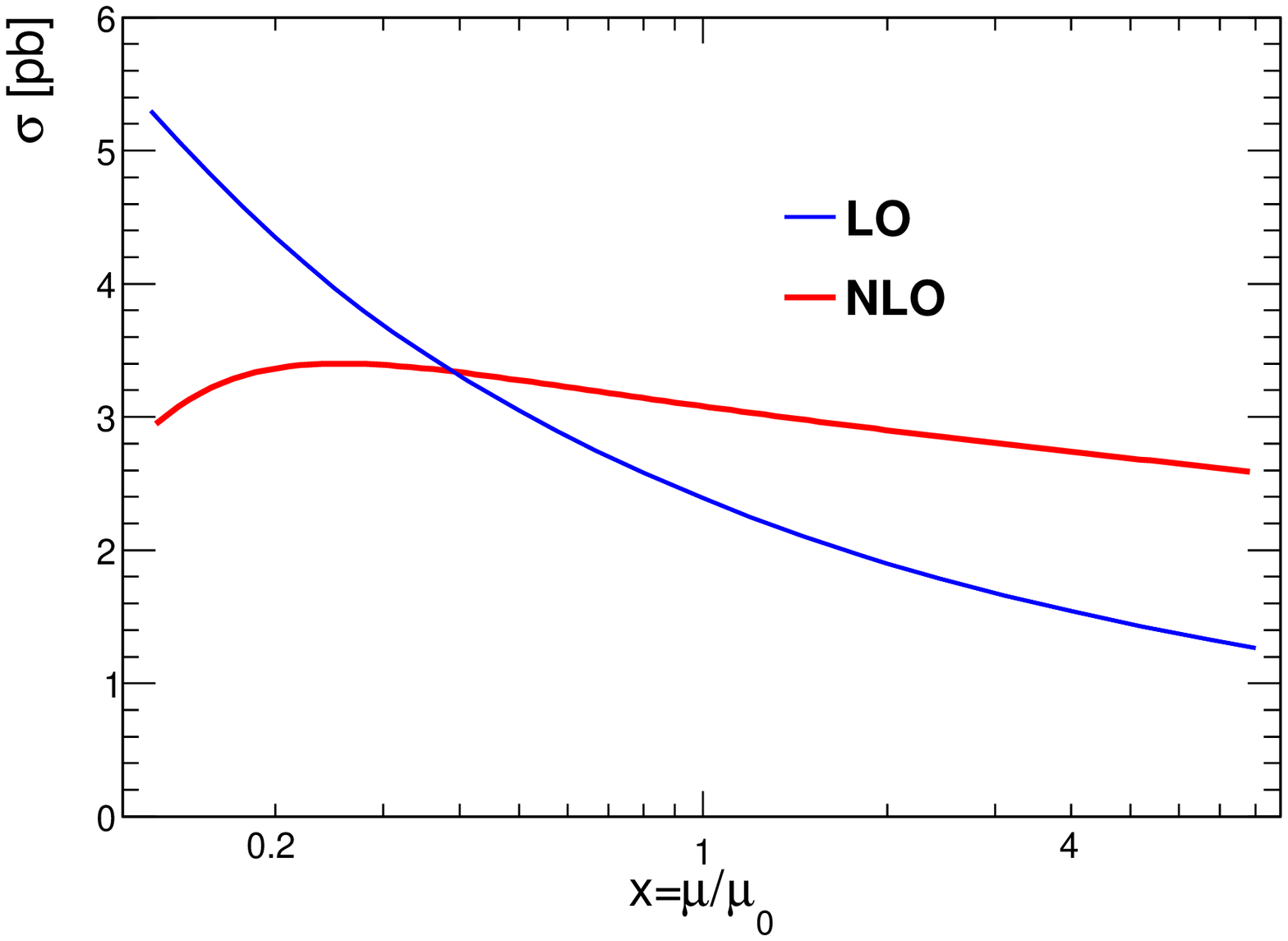} 
\caption{Scale dependence of the total cross section at LO and NLO with $x= \mu/\mu_0$.}
\label{Fig:scalevar}
\end{center}
\end{figure}

The numerical results presented in the following have been calculated at a
center-of-mass energy  of $\sqrt{s}=8$\,TeV.
For the  jet clustering we used an anti-$k_T$ algorithm~\cite{Cacciari:2008gp} with a cone size
of $R_j=0.5$ provided by 
the  FastJet package \cite{fastjet}.
We used the CT10 set of parton distributions~\cite{Lai:2010vv} 
as contained in the CT10.LHgrid set of the LHAPDF library~\cite{Whalley:2005nh} for both LO and NLO
contributions and worked with $N_F=5$ massless 
quark flavors.
The following kinematic cuts, based on recent experimental studies~\cite{cmsdiphoton}, have been applied:
$p_T^{\rm{jet}}>30\mbox{ GeV}, p_T^{\gamma,1}>40 \mbox{~GeV},p_T^{\gamma,2}>25\mbox{~GeV},|\eta^{\gamma}| \leq 2.5, 
|\eta^{j}| \leq 4.7, R_{\gamma ,j} > 0.5, R_{\gamma, \gamma} >0.45.$
For the photon isolation we restricted ourselves to the Frixione isolation criterion with 
$R=0.4, n=1$ and $\epsilon=0.05$. A more detailed analysis with different isolation criteria and various
isolation parameters will be presented elsewhere. The scales have been chosen as in section \ref{sec:1j}.
Figure~\ref{Fig:scalevar} shows the scale variation around the central value. A clear reduction of
the scale dependence can be observed by adding the next-to-leading order contributions with a moderate 
K-factor of $\sim1.3$ for the central scale.
The effects of the next-to-leading order corrections in differential distributions can be substantial
even if the effects on the total cross section are rather moderate. This is in particular true, if at NLO,
regions in phase space open up that are kinematically not allow at leading order.
\begin{figure}[t]
\begin{center}
\includegraphics[width=7.5cm]{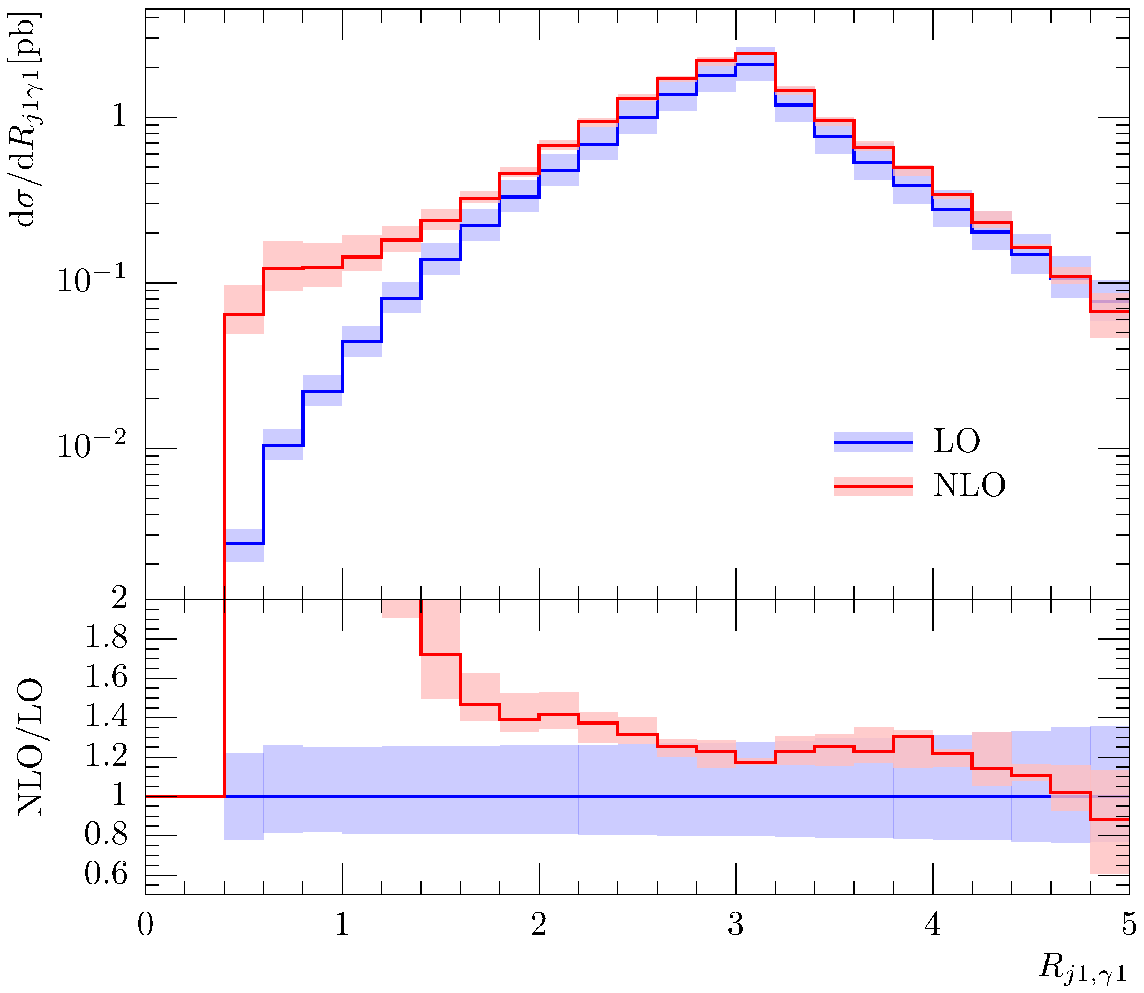} 
\caption{$R$-separation $R(j_1,\gamma_1)$ between the hardest jet and the hardest photon.}
\label{Fig:Rj1a1}
\end{center}
\end{figure}
Figure \ref{Fig:Rj1a1} shows the $R$-separation between the hardest jet and the hardest photon.
The colored bands denote the scale uncertainty when varying around the central scale by a factor of two.
For small values of $R$, the NLO corrections lead to a strong distortion of the shape. This can be understood
in terms of kinematically allowed regions. At LO, the pair of hard photon and hard jet being close in $R$-space
would have to be counterbalanced with the
soft photon and the soft jet. This is kinematically impossible at LO but 
appears at NLO due to the additional radiation. 

\section{Conclusions}
We have presented the next-to-leading order QCD corrections to the production of a photon pair in
association with one and two jets. This class of processes is important for a reliable prediction
of the Standard Model background to the $H\to\gamma\gamma$ signal. In the one jet case we have implemented
and compared two types of photon isolation criteria, the fixed cone isolation and Frixione isolation.
For both cases we find large NLO corrections and for the inclusive case a strong dependence on the
scale. The scale dependence can be effectively reduced by imposing a veto on a second jet. In general
we find a strong dependence on isolation types and parameters, however it can be observed that for 
tight isolation parameters, the results obtained with different isolation types approach each other.
We also calculated the first NLO corrections to the process of diphoton plus two jet production.
For the photon isolation we used Frixione isolation and found moderate NLO corrections to the total
cross section. The effects on differential distributions depend on the observable and can lead to
strong distortions of the shape.

\section*{Acknowledgments}
We thank the other members of the GoSam collaboration for useful discussions.
N.G. and G.H. want to thank the University of Zurich for kind hospitality
while parts of this project were carried out.
This work was supported in part by the Schweizer Nationalfonds under grant
200020-138206, and by the Research Executive Agency (REA) of the
European Union under the Grant Agreement number PITN-GA-2010-264564
(LHCPhenoNet). We acknowledge use of the computing resources of the Rechenzentrum Garching.

\end{document}